\documentstyle[preprint,aps]{revtex}
\begin{document}
\draft
\preprint{SNUTP 95-99}
\title{Monopoles in Weinberg-Salam Model}
\author{Y. M. Cho}
\address{Department of Physics, Seoul National University, Seoul 151, Korea}
\author{D. Maison}
\address{Max-Planck-Institut f\"ur Physik, Werner-Heisenberg-Institut,
          F\"ohringer Ring 6\\ 80805 Munich, Germany}

\maketitle

\begin{abstract}
We present a new type of spherically symmetric monopole and dyon solutions
with the magnetic charge $ 4\pi/e$ in the standard Weinberg-Salam model.
The monopole (and dyon) could be interpreted 
as a non-trivial hybrid between the 
abelian Dirac monopole and non-abelian 't Hooft-Polyakov monopole (with 
an electric charge).
We discuss the possible physical implications of the electroweak dyon.
\end{abstract}

\newpage
Ever since Dirac \cite{Dirac} has generalized the Maxwell's theory with his
magnetic monopole, the monopoles have been a subject of extensive
studies.
The Abelian monopole has been generalized to the non-Abelian 
gauge theory by Wu and Yang \cite{Wu} who constructed 
a non-Abelian monopole solution 
in the pure $SU(2)$ gauge theory,
and by 't Hooft and Polyakov \cite{Hooft} who  have shown that 
the $SU(2)$ gauge theory
allows a finite energy monopole solution as a topological soliton in
the presence of a triplet scalar source. 
In the interesting case of the electroweak theory of Weinberg and
Salam \cite{Wein}, however, it has generally been believed that there
exists no topological monopole of physical interest.
The basis for this ``non-existence theorem'' is, of course, that with
the spontaneous symmetry breaking the quotient space $SU(2) \times
U(1)/U(1)_{\rm em}$ allows no non-trivial second homotopy.
This has led many people  to conclude that 
there is no topological structure in
the Weinberg-Salam model which can accommodate a magnetic monopole.
The purpose of this letter is to show that this conclusion is
premature.
In the following {\it we establish the existence of a new type of 
monopole and dyon solutions in the standard Weinberg-Salam
model, and clarify the topological origin of the magnetic charge}.
Clearly  the new monopole and dyon will have important implications 
in the phenomenology of the electroweak theory, 
which can make them very interesting from the physical point of view.

Before we construct the monopole we must understand how one
can circumvent the non-existence theorem in the Weinberg-Salam model
and obtain the desired solutions.
For this it is important to realize that, 
{\it with the extra hypercharge $U(1)$
degrees of freedom, the standard Weinberg-Salam model could
be viewed as a gauged $CP^1$ model in which the (normalized) Higgs
doublet plays the role of the $CP^1$ field}. 
Viewed as a $CP^1$ doublet the Higgs field can now admit a topologically
non-trivial configuration whose second homotopy is given
by $\pi_2(CP^1)=Z$. This clears the way for a genuine topological
monopole in the Weinberg-Salam model which can be described by a completely
regular $SU(2)$ potential.

To construct the desired solutions we start with the Lagrangian which
describes (the bosonic sector of) the standard Weinberg-Salam model 
\begin{eqnarray*}
{\cal L} = -{|\hat{D}_{\mu}\mbox{\boldmath $\phi$}|}^2 -
\frac{\lambda}{2}\left( \mbox{\boldmath $\phi$}^\dagger
\mbox{\boldmath $\phi$}-\frac{\mu^2}{\lambda}\right)^2 -
\frac{1}{4} (\mbox{\boldmath $ F$}_{\mu\nu})^2 - \frac{1}{4}(G_{\mu\nu})^2,
\end{eqnarray*}
\begin{eqnarray}
\hat{D}_{\mu} \mbox{\boldmath $\mbox{\boldmath ${\phi}$}$}
&=& \left (\partial_{\mu} - i
\frac{g}{2}\mbox{\boldmath $\tau$}\!\!\cdot\!\mbox{\boldmath $A$}_{\mu}
 - i
\frac{g'}{2} B_{\mu}\right) \mbox{\boldmath $\phi$}\label{eq:Lag}\\
                      &=& \left (D_{\mu}- i \frac{g'}{2}
                        B_{\mu}\right) \mbox{\boldmath $\phi$},\nonumber
\end{eqnarray}
where ${\bf \mbox{\boldmath $\phi$}}$ is the Higgs doublet, 
$\mbox{\boldmath $F$}_{\mu\nu}$ and
$G_{\mu\nu}$ are the gauge fields of $SU(2)$ and $U(1)$ with the
potentials $\mbox{\boldmath $A$}_{\mu}$ and $B_{\mu}$, and $g$ and $g'$
are the corresponding coupling constants.
Notice that $D_{\mu}$ describes the covariant derivative  of the $SU(2)$
subgroup only. From (1) one has the following equations of motion
\begin{eqnarray}
&&\hat{D}_{\mu}(\hat{D}_{\mu}\mbox{\boldmath $\phi$}) =
   \lambda \left(\mbox{\boldmath ${\phi}$}^{\dagger} 
\mbox{\boldmath ${\phi}$}- \frac {\mu^2}{\lambda} \right) 
\mbox{\boldmath ${\phi}$},\nonumber\\
&&D_{\mu} \mbox{\boldmath $F$}_{\mu\nu} = -\mbox{\boldmath $j$}_{\nu} = i
\frac{g}{2} \left[ \mbox{\boldmath ${\phi}$}^{\dagger} \mbox{\boldmath $\tau$}
  (\hat{D}_{\nu}\mbox{\boldmath ${\phi}$}) 
- (\hat{D}_{\nu} \mbox{\boldmath ${\phi}$})^{\dagger}
  \mbox{\boldmath $\tau$} \mbox{\boldmath ${\phi}$}\right] , \label{eq:Eq}\\
&&\partial_{\mu} G_{\mu\nu} = -k_{\nu} = i \frac{g'}{2} \left[
  \mbox{\boldmath ${\phi}$}^{\dagger} (\hat{D}_{\nu} 
\mbox{\boldmath ${\phi}$}) 
- (\hat{D}_{\nu} \mbox{\boldmath ${\phi}$})^{\dagger}
  \mbox{\boldmath ${\phi}$}\right]. \nonumber
\end{eqnarray}
Notice that with
\begin{eqnarray}
\mbox{\boldmath ${\phi}$}&=& \frac{1}{\sqrt{2}}\rho \, \xi \hspace{10mm}
( \rho^2 = 2 \mbox{\boldmath ${\phi}$}^{\dagger}
\mbox{\boldmath ${\phi}$},\hspace{5mm}   \xi^{\dagger}\xi=1) ,\nonumber\\
{\mbox{\boldmath $\hat{\phi}$}} 
&=& \xi^{\dagger} \mbox{\boldmath $\tau$} \xi , \nonumber\\
A_{\mu} &=& {\mbox{\boldmath $\hat{\phi}$}}
\! \cdot\!  \mbox{\boldmath $A$}_{\mu}, 
\label{eq:Def} \\
C_{\mu} &=&  i \xi^{\dagger}  \partial_{\mu}  \xi, \nonumber
\end{eqnarray}
one has 
\begin{eqnarray*}
\mbox{\boldmath $ j$}_{\mu} &=&- \frac{g \rho^2}{2} \left[\frac{g}{2}
  \mbox{\boldmath $A$}_{\mu} + \left(\frac{g'}{2} B_{\mu} + C_{\mu}\right)
  {\mbox{\boldmath $\hat{\phi}$}} + \frac{1}{2} 
{\mbox{\boldmath $\hat{\phi}$}} \times \partial_{\mu} 
{\mbox{\boldmath $\hat{\phi}$}}
\right] ,    \\
k_{\mu} &=& -\frac{ g'\rho^2}{2} 
          \left(\frac{g}{2}A_{\mu} + \frac{g'}{2}B_{\mu} +
C_{\mu}\right) = \frac{g'}{g}({\mbox{\boldmath $\hat{\phi}$}}\!\cdot\! 
\mbox{\boldmath $j$}_{\mu})  .
\end{eqnarray*}
Now we choose the following static spherically symmetric ansatz
\begin{eqnarray*}
\rho &=& \rho(r),\\[4mm]
\xi &=& i \left(\begin{array}{cc}
\sin (\theta/2)\,\, e^{-i\varphi}\\
- \cos(\theta/2)
\end{array} \right), 
\hspace{5mm}{\mbox{\boldmath $\hat{\phi}$}} = \xi^{\dagger}
\mbox{\boldmath $\tau$} \xi = - \hat{r},
\end{eqnarray*}
\begin{eqnarray}
\mbox{\boldmath $A$}_{\mu} &=& \frac{1}{g} A(r)
\partial_{\mu}t{\mbox{\boldmath $\hat{\phi}$}} 
+ \frac{1}{g}(f(r)-1) {\mbox{\boldmath $\hat{\phi}$}} \times
\partial_{\mu} {\mbox{\boldmath $\hat{\phi}$}} \label{eq:Ansa},\\[4mm]
B_{\mu} &=& - \frac{1}{g'} B(r) \partial_{\mu}t -
\frac{1}{g'}(1-\cos\theta) \partial_{\mu} \varphi, \nonumber
\end{eqnarray}
where $(t,r, \theta, \varphi)$ are the polar coordinates.
Notice that the apparent string
singularity along the negative z-axis in $\xi$ and $B_{\mu}$ is a pure
gauge artifact which can easily be removed with a hypercharge $U(1)$
gauge transformation. Indeed 
one can easily exociate the string by making the hypercharge $U(1)$ 
bundle non-trivial \cite{Wu}. So {\it the above ansatz describes a most
general spherically symmetric ansatz of a $SU(2) \times U(1)$ dyon}.
Here we emphasize the importance of the non-trivial $U(1)$ 
degrees of freedom to
make the ansatz spherically symmetric. Without the extra $U(1)$ the
Higgs doublet does not allow a spherically symmetric ansatz.
This is because the spherical symmetry for 
the gauge field involves the embedding
of the radial isotropy group $SO(2)$ into the gauge group 
that requires the Higgs field to be
invariant under the $U(1)$ subgroup of $SU(2)$. This is possible
with a Higgs triplet, 
but not with a Higgs doublet \cite{Forg}. In fact, in the
absence of the hypercharge $U(1)$ degrees of freedom, the above ansatz
describes the $SU(2)$ sphaleron which is not spherically
symmetric \cite{Dashen}. The situation changes with the 
inclusion of the extra hypercharge 
$U(1)$ in the standard model, which can compensate the action of
the $U(1)$ subgroup of $SU(2)$ on the Higgs field. 

With the spherically symmetric  ansatz (2)
is reduced to  
\begin{eqnarray}
&&\ddot{f} - \frac{f^2-1}{r^2}f = 
              \left(\frac{g^2}{4}\rho^2 - A^2\right)f, \nonumber
\\[4mm]
&&\ddot{\rho} + \frac{2}{r} \dot{\rho} - \frac{1}{2} \frac{f^2}{r^2}\rho
  =- \frac{1}{4}(A-B)^2\rho + \lambda\left(\frac{\rho^2}{2} -
   \frac{\mu^2}{\lambda}\right)\rho \nonumber, \\[4mm]
&&\ddot{A} + \frac{2}{r}\dot{A} -2\frac{f^2}{r^2}A = \frac{g^2}{4}
   \rho^2(A-B), \label{eq:Spher}\\[4mm]
&&\ddot{B} + \frac{2}{r} \dot{B} =  \frac{g'^2}{4} \rho^2 (B-A). \nonumber  
\end{eqnarray}
At this point one may wish to compare our dyon with that of 
Julia and Zee \cite{Julia}, which is obtained from the familiar 
Lagrangian 
\begin{eqnarray}
\label{eq:Julia}
 {\cal L'}=-\frac{1}{2}|D_\mu\Phi|^2
-\frac{\lambda}{4}\left(\Phi^2-\frac{\mu^2}{\lambda}\right)^2
-\frac{1}{4}(\mbox{\boldmath $F$}_{\mu\nu})^2 ,
\end{eqnarray}
where $\Phi$ is the Higgs triplet. With 
\begin{eqnarray}
\label{eq:Zee}
 && \Phi=-\rho(r)\hat{r} \nonumber,\\
&& \mbox{\boldmath $A$}_\mu=-\frac{1}{g}A(r)\partial_\mu t\,\,\hat{r}
  + \frac{1}{g}(f(r)-1)\hat{r}\times\partial_\mu \hat{r},
\end{eqnarray}
the Julia-Zee dyon is described by
\begin{eqnarray}
\label{eq:JuliaZee}
&&\ddot{f}- \frac{f^2-1}{r^2}f=\left(g^2\rho^2-A^2\right)f, \nonumber\\
&&\ddot{\rho}+\frac{2}{r}\dot{\rho} - 2\frac{f^2}{r^2}\rho =
     \lambda \left(\rho^2 - \frac{\mu^2}{\lambda} \right)\rho , \\  
&&\ddot{A}+\frac{2}{r}\dot{A} -2\frac{f^2}{r^2} A=0 . \nonumber
\end{eqnarray}
This shows that there is a remarkable similarity between the two
dyons.
A closer comparison between the two dyons will be discussed soon.

To integrate (\ref{eq:Spher}) one may choose the following boundary condition
\begin{eqnarray}
\label{eq:Bound}
\begin{array}{llll}
f(0)=1, & \rho(0)=0,&A(0)=0, & B(0)=b_0, \\
f(\infty)=0, & \rho(\infty)= \rho_0 , &
A(\infty)=A_0,&B(\infty)=B_0 ,
\end{array} 
\end{eqnarray}
which guarantees the regularity of the solutions in the $SU(2)$ sector.
With this one can easily show that near the origin one must have
\begin{eqnarray}
\label{eq:Origin}
&& f \simeq 1+ \alpha_1  r^2+\cdots, \nonumber \\
&& \rho \simeq \beta_1 r^\delta +\cdots, \nonumber \\
&& A\simeq a_1 r+\cdots,\\
&& B\simeq b_0 +b_1 r+\cdots, 
          \nonumber 
\end{eqnarray}
where $\delta =(-1+\sqrt{3})/2$. 
On the other hand asymptotically one must have 
\begin{eqnarray}
\label{eq:Infty}
&& f \simeq  f_1 \exp(-\kappa  r)+\cdots,\nonumber\\
&& \rho \simeq \rho_0 +\rho_1\frac{\exp(-\sqrt{2}\mu r)}{r}
  +\cdots, \nonumber \\
&& A \simeq A_0 +\frac{A_1}{r}+ \cdots , \\
&& B \simeq A+B_1 \frac{\exp(-\nu r)}{r}+\cdots, \nonumber 
\end{eqnarray}
where $\rho_0=\sqrt{2/\lambda}\mu$, 
$\kappa=\sqrt{(g\rho_0)^2/4 -A_0^2}$,
and $\nu=\sqrt{(g^2 +g'^2)}\rho_0/2$. 
Notice  that asymptotically $B(r)$ must approach to $A(r)$ with an 
exponential damping.

To determine the possible electric and magnetic charge of the desired
solutions we now perform the following gauge transformation
on (\ref{eq:Ansa})
\begin{eqnarray}
\xi \longrightarrow U \xi = \left(\begin{array}{cc}
0 \\ 1
\end{array} \right) ,
\label{eq:Gauge}
\end{eqnarray}
\begin{eqnarray}
 U=-i\left( \begin{array}{cc}
        \cos (\theta/2)& \sin(\theta/2)e^{-i\varphi} \\
        \sin(\theta/2) e^{i\varphi} & -\cos(\theta/2)
\end{array}
\right),\nonumber
\end{eqnarray}
and find that in this unitary  gauge
\begin{eqnarray}
\mbox{ \boldmath $A$}_\mu \longrightarrow
\frac{1}{g} 
\left(
\begin{array}{c}
(\sin\varphi\partial_\mu\theta 
     +\sin\theta\cos\varphi \partial_\mu\varphi )f(r)\\
(-\cos\varphi\partial_\mu \theta
     +\sin\theta\sin\varphi\partial_\mu\varphi )f(r)\\
-A(r)\partial_\mu t -(1-\cos\theta)\partial_\mu\varphi
\end{array}
\right).\label{eq:Unitary}
\end{eqnarray}
In particular we have 
\begin{eqnarray}
A^3_{\mu} &=& - \frac{1}{g} A(r) \partial_{\mu}t -
\frac{1}{g}(1-\cos\theta) \partial_{\mu} \varphi, \nonumber\\
B_{\mu} &=& - \frac{1}{g'}B(r) \partial_{\mu}t -
\frac{1}{g'}(1-\cos\theta) \partial_{\mu} \varphi. \label{eq:Compo}
\end{eqnarray}
So expressing the electromagnetic potential ${\cal A}_\mu$ and the 
neutral potential ${\cal Z}_\mu$ with the Weinberg angle $\theta_{\rm w}$
\begin{eqnarray}
\left( \begin{array}{cc}
{\cal A}_{\mu} \\ {\cal Z}_{\mu} 
\end{array} \right)
&=& \left(\begin{array}{cc}
\cos\theta_{\rm w} & \sin\theta_{\rm w}\\
-\sin\theta_{\rm w} & \cos\theta_{\rm w}
\end{array} \right)
\left( \begin{array}{cc}
B_{\mu} \\ A^3_{\mu}
\end{array} \right) \nonumber\\
&=& \frac{1}{\sqrt{g^2 + g'^2}}
\left(\begin{array}{cc}
g & g' \\ -g' & g
\end{array} \right)
\left( \begin{array}{cc}
B_{\mu} \\ A^3_{\mu}
\end{array} \right) , \label{eq:Wein}
\end{eqnarray}
we have 
\begin{eqnarray}
\label{eq:Em}
{\cal A}_{\mu} &=& - e \left( \frac{1}{g^2}A +
    \frac{1}{g'^2}B \right) \partial_{\mu}t -
  \frac{1}{e}(1-\cos\theta) \partial_{\mu}
 \varphi,  \nonumber \\
{\cal Z}_{\mu} &=& \frac{e}{gg'}(B-A) \partial_{\mu}t , 
\end{eqnarray}
where $e$ is the electric charge 
$$
e=\frac{gg'}{\sqrt{g^2+g'^2}}.
$$ 
From this we conclude that the desired solutions should carry 
the following electromagnetic charges 
\begin{eqnarray}
\label{eq:Charge}
q_e &=& {4\pi e}\left[
r^2 \left.\left(\frac{1}{g^2}\dot{A}+\frac{1}{g'^2}\dot{B} 
\right)\right]        \right|_{r=\infty}
 =\frac{4\pi}{e} A_1 \nonumber \\
 &=&\frac{8\pi}{e}\sin^2\theta_{\rm w}\int^\infty_0 f^2 A dr , \\
q_m &=& \frac{4\pi}{e}. \nonumber  
\end{eqnarray}
Also, from the asymptotic condition (\ref{eq:Infty})
we conclude that our solutions should not carry any neutral charge,
\begin{eqnarray}
\label{eq:Neutral}
&&{\cal Z}_e =-\frac{4\pi e}{gg'}\left.\left[ r^2 (\dot{B}-\dot{A})\right]
\right|_{r=\infty}
=0,\nonumber \\ 
&&{\cal Z}_m = 0,
\end{eqnarray}
which is what one would have expected.

With this  one may now try to find out the desired solutions.
Although it appears that (\ref{eq:Spher}) 
does not allow a solution which can be expressed 
in terms of elementary functions, 
we find that we can integrate it numerically when $\kappa$ is positive. 
The monopole solution with $A=B=0$ is 
shown in Fig.1, and a typical dyon solution is shown in Fig.2.
As expected our solutions indeed 
look very similar to the well-known
Prasad-Sommerfield solutions of the Julia-Zee dyon \cite{Julia}.
But, of course, there is a crucial difference. 
{\it The new feature here is that our dyon now has a non-trivial
$B-A$, which represents the neutral $Z$ boson content 
of the dyon as shown by (16)}.
To understand the behavior of the solutions, remember
that the mass of the $W$ and $Z$ bosons are given by 
$M_W=g\rho_0/2$ and $M_Z=\sqrt{g^2+g'^2}\rho_0/2$, and the 
mass of Higgs boson is given by $M_H=\sqrt{2}\mu$.
So our result confirms that $\sqrt{(M_W)^2-(A_0)^2}$ 
and $M_H$ determines the exponential
damping of $f$ and $\rho$, and $M_Z$ determines the 
exponential damping of $B-A$, to their vacuum expectation values
asymptotically.
These are exactly what one would have expected.

The canonical energy of the dyon is given by 
\begin{eqnarray}
\label{eq:Energy}
&&E=E_0+E_1,\nonumber\\
&&E_0=\frac{2\pi}{g'^2}M_W\int_0^\infty \frac{dx}{x^2},\\
&&E_1= \frac{4\pi}{g^2}M_W   \int_0^\infty dx 
\left[ \left( \frac{df}{dx}\right)^2 
+\frac{(f^2-1)^2}{2 x^2}
+\frac{2x^2}{g^2} \left(\frac{d}{dx}\frac{A}{\rho_0}\right)^2
+\frac{4f^2}{g^2}\left( \frac{A}{\rho_0}\right)^2
+\frac{2x^2}{g'^2}\left(\frac{d}{dx}\frac{B}{\rho_0}\right)^2
      \right.\nonumber \\
&&\hspace{40mm}
+ {2x^2} \left(\frac{d}{dx}\frac{\rho}{\rho_0}\right)^2
+f^2\left(\frac{\rho}{\rho_0}\right)^2 
+\frac{2x^2}{g^2} \left(\frac{\rho}{\rho_0}\right)^2
 \left(\frac{A}{\rho_0}-\frac{B}{\rho_0}\right)^2 \nonumber \\
&&\hspace{40mm}\left.
+\frac{2\lambda x^2}{g^2}\left(
\left( \frac{\rho}{\rho_0}\right)^2 - 1\right)^2
\right],\nonumber
\end{eqnarray}
where  $x= M_W r$ is a dimensionless variable.
So the classical energy of the dyon is
made of two parts, the infinite part $E_0$ which solely comes from the
point-like hypercharge magnetic monopole and the finite part $E_1$
which comes from the rest.
One might worry about 
the infinite energy  $E_0$ of the dyon. 
But from the physical point of view this need
not be  a serious drawback. The infinite part is still controlled by the
weak energy scale $M_W$, and could easily be made finite by
embedding the $SU(2)\times U(1)$ to a larger group \cite{Dokos}. 
It could also be made finite with the 
introduction of the gravitational interaction \cite{Bais}. 
Furthermore it could be treated as
the ``vacuum" energy when one quantizes the classical dyon.
So one could easily subtract or factorize out the infinite part and
obtain a finite result in the physical applications of the dyon.

To clarify the topological origin of our dyon it is
important to understand
the similarity between our dyon and the Julia-Zee dyon. For this
notice that the Lagrangian (\ref{eq:Julia})  with
the Higgs triplet scalar source $\Phi $, which allows the Julia-Zee dyon,
can be viewed to describe  a $CP^1$ gauge theory. 
Indeed with the identification 
\begin{eqnarray}
\label{eq:Tri}
\Phi = \rho {\mbox{\boldmath $\hat{\phi}$}} ,\hspace{10mm}
{\mbox{\boldmath $\hat{\phi}$}} = \xi^{\dagger}
\mbox{\boldmath $\tau$} \xi ,
\end{eqnarray}
one can easily show that the Lagrangian (\ref{eq:Julia}) 
can be written as 
\begin{eqnarray}
\label{eq:lagran}
{\cal L'} 
= - \frac{1}{2} {(\partial_{\mu} \rho)}^2 
-\frac{\lambda}{4}\left(\rho^2 -\frac{\mu^2}{\lambda}\right)^2 
 - 2{\rho}^2
\left({|D_{\mu}\xi|}^2 - {|\xi^{\dagger}D_{\mu}\xi|}^2\right)
- \frac{1}{4}( \mbox{\boldmath $F$}_{\mu\nu})^2.
\end{eqnarray}
On the other hand the Lagrangian (\ref{eq:Lag}) with (\ref{eq:Def}) 
can also be written as
\begin{eqnarray}
{\cal L} &=& -\frac{1}{2}{(\partial_{\mu}\rho)}^2 
-\frac{\lambda}{2}\left(\frac{1}{2}\rho^2-\frac{\mu^2}{\lambda}\right)^2
  - \frac{1}{2}{\rho}^2
{|\hat{D}_{\mu} \xi |}^2 -  \frac{1}{4}(\mbox{\boldmath $F$}_{\mu\nu})^2
- \frac{1}{4} (G_{\mu\nu})^2 \nonumber\\
&=& -\frac{1}{2}{(\partial_{\mu} \rho)}^2 
-\frac{\lambda}{2}\left(\frac{1}{2}\rho^2-\frac{\mu^2}{\lambda}\right)^2 
 - \frac{1}{2}{\rho}^2 \left({|D_{\mu}\xi|}^2
- {|\xi^{\dagger}D_{\mu}\xi|}^2\right) \label{eq:Dyon}\\
&&+
  \frac{1}{2}{\rho}^2{\left(\xi^{\dagger}D_{\mu}\xi 
      - i \frac{g'}{2}B_{\mu}\right)}{}^2 -
  \frac{1}{4} (\mbox{\boldmath $F$}_{\mu\nu})^2 
- \frac{1}{4}( G_{\mu\nu})^2. \nonumber
\end{eqnarray}
Now a simple  comparison between (\ref{eq:lagran}) 
and (\ref{eq:Dyon}) tells that they are almost identical. 
Indeed the only difference between the two Lagrangians 
(other than the constant normalization coefficients) 
is the interaction of the $U(1)$ gauge field to  the
other fields through the term $\rho ^2{(\xi^{\dagger} D_{\mu} \xi -
  i(g'/2) B_{\mu})}^2$ in (\ref{eq:Dyon}), which is invariant
under the $U(1)$ gauge transformation of the $CP^1$ field $\xi$.
This shows that the Weinberg-Salam model can also be
viewed as a $CP^1$ gauge theory, in which the Higgs 
doublet can easily accomodate a non-trivial topology
described by $\pi_2(CP^1)=Z$. This implies that, with a
judicious choice of an ansatz, the Julia-Zee dyon  
could be extended to a $SU(2) \times U(1)$ electroweak dyon. More
importantly {\it this shows that our dyon has 
exactly the same topological origin
as the Julia-Zee dyon, $\pi_2(CP^1)=Z$}. The new feature
here is that when $\pi_2(CP^1)$ become non-trivial 
the hypercharge $U(1)$ should also become non-trivial,
which is due to the $SU(2) \times U(1)$ invariant interaction.

To summarize, we have presented a new type of 
spherically symmetric electroweak monopole (and dyon)
which can be interpreted as a hybrid between the Abelian Dirac monopole
and the non-Abelian 't Hooft-Polyakov monopole (with an electric charge)
in the $SU(2)\times U(1)$
gauge theory of Weinberg-Salam. Obviously the monopole must be stable because
the magnetic charge cannot evaporate. Of course,
the electric charge of our dyon remains
a free parameter at the classical level. But, just like the Julia-Zee dyon,
the electric charge will be quantized after the quantization of the
classical dyon.
We close with the following remarks:\\
1) As we have emphasized, our monopole 
can be viewed as a electroweak generalization  of the Dirac monopole. 
As obviously it can be also viewed as a electroweak generalization of 
the 't Hooft-Polyakov monopole. 
Nevertheless there is an important difference between these and our monopole. 
For the known monopoles 
(of Dirac and 't Hooft-Polyakov) it is well-known
that the magnetic charge $q_m$ obeys the Dirac
quantization condition $q_m=2\pi n/e$ ($n$ ; integers), 
where $e$ is the minimum electric charge of the theory. 
In contrast our spherically symmetric  monopole
carries the minimum magnetic charge $4\pi/e$. 
This  
suggests that the magnetic charge of the electroweak monopole could obey the
Schwinger quantization condition $q_m=4\pi n/e$, rather than the Dirac 
condition.

\noindent 2) The fact that our dyon is a electroweak 
dyon suggests that it should
be taken seriously as a realistic object. 
Assuming its existence, it could
have important  phenomenological consequences. 
For example in the leptonic sector 
the possibility of the
helicity changing scattering process at the electroweak scale, or   
the fermionic zero-mode (zero-energy bound state)
in the presence of the monopole, should be carefully re-analyzed.
 Furthermore in the hadronic
sector the monopole catalysis of  the proton decay through the Callan-Rubakov
effect \cite{Callan}, which should  
depend on how one embeds the electroweak $SU(2)\times U(1)$ 
to a larger group to construct the grand unified theory,
 should also be re-examined. 

\noindent 3) Finally it must be emphasized that our dyon is different from
the one that Nambu and others have discussed before \cite{Nambu}, which 
describes a monopole connected to an anti-monopole 
by a neutral but real (physical)
string. So the Nambu's monopole is not spherically symmetric, and
does not approach
to the vacuum configuration asymtotically. In contrast
our dyon describes a genuine isolated spherically symmetric dyon
which is not attached to any physical string, and approaches to the vacuum
configuration asymtotically. Topologically, the difference
can be traced to the fact that  in the Nambu's case 
the hypercharge $U(1)$ bundle is trivial but in our case 
the $U(1)$ bundle becomes non-trivial.

There are many other issues, both mathematical and
physical, which need to be addressed in  more detail
concerning the new dyon.
We will discuss these issues  in detail in a
separate paper.
\acknowledgements

One of us (YMC) thanks G. 't Hooft for the illuminating discussions.
The work is supported in part by the Ministry of Education and by the Korean
Science and Engineering Foundation.
\newpage
\newpage
\centerline{\bf Figure Captions}

\noindent {\bf Fig.1.} The electroweak monopole solution, where we have chosen  
$\sin^2\theta_{\rm w}=0.2325$ and $M_H/M_W=1$ 
($\lambda/g^2=1/4$).
The plot shows $f(r)$ and  $\rho(r)$ 
as functions of dimensionless variable $x=M_W r$.

\setlength{\unitlength}{0.240900pt}
\ifx\plotpoint\undefined\newsavebox{\plotpoint}\fi
\sbox{\plotpoint}{\rule[-0.200pt]{0.400pt}{0.400pt}}%

\vskip 2em

\vskip 1em
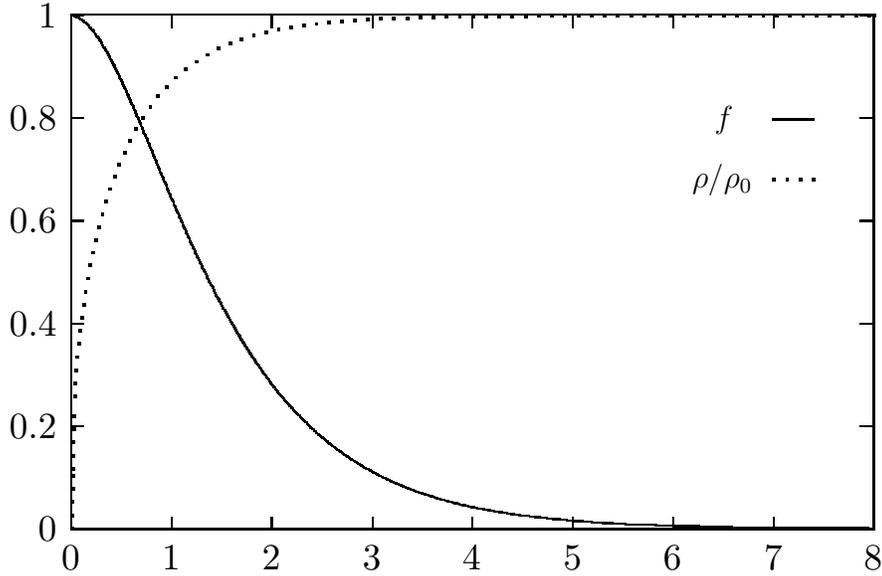
\noindent {\bf Fig.2.} A typical  electroweak dyon solution, where
we have chosen $A_0=M_W /2$. The plot shows
$f(r)$, $\rho(r)$, $A(r)$, and $Z(r)=B(r)-A(r)$ as functions of dimensionless
variable $x=M_W r$.  
\vskip 2em

\setlength{\unitlength}{0.240900pt}
\ifx\plotpoint\undefined\newsavebox{\plotpoint}\fi
\sbox{\plotpoint}{\rule[-0.200pt]{0.400pt}{0.400pt}}%


\end{document}